\documentclass[aps,prl,reprint, amsmath, amssymb,superscriptaddress]{revtex4-2}
 
\usepackage{bm}
\usepackage[retainorgcmds]{IEEEtrantools}
\usepackage{graphicx}
\usepackage{mathrsfs}
\usepackage{amsmath}
\usepackage{amsfonts}
\usepackage{amssymb}
\usepackage{color}
\usepackage{times,txfonts}
\usepackage{nicefrac}
\usepackage{ragged2e}
\usepackage{tikz}
\usepackage{subfigure}
\usepackage{braket}

\usepackage[colorlinks=true,linkcolor=blue,urlcolor=blue,citecolor=blue,pdfusetitle]{hyperref}

\usepackage{blindtext}

\definecolor{cadmiumgreen}{HTML}{097969}

\begin{document}



\title{Two-point measurement energy statistics from particle scattering}


\author{Samuel L. Jacob}
\email{sajacob@tcd.ie}
\affiliation{Department of Physics, Trinity College Dublin, Dublin 2, Ireland}

\date{\today}
\author{Gabriel T. Landi}
\affiliation{Department of Physics and Astronomy, University of Rochester, Rochester, New York 14627, USA}
\email{gabriel.landi@rochester.edu}

\author{Massimiliano Esposito}
\email{massimiliano.esposito@uni.lu}
\affiliation{Complex Systems and Statistical Mechanics, Physics and Materials Science Research Unit, University of Luxembourg, L-1511 Luxembourg, G.D. Luxembourg}

\author{Felipe Barra}
\email{fbarra@dfi.uchile.cl }
\affiliation{Departamento de F\'isica, Facultad de Ciencias F\'isicas y Matem\'aticas, Universidad de Chile, 837.0415 Santiago, Chile}

\begin{abstract}

We show that the energy statistics resulting from a two-point measurement of an isolated quantum system subject to a time-dependent driving protocol can be probed by subjecting the same system to a collision with a suitably prepared incoming particle. This means that the particle acts both as an external drive and as an energy measurement device for the quantum system and that energy fluctuations can be defined within a fully autonomous setting.

\end{abstract}

\maketitle{}

\emph{Introduction. ---} Fluctuations are ubiquitous at the micro-scale \cite{Jarzynski2011,Seifert2012}, but properly defining quantum fluctuations is not always an easy task. This is particularly true out-of-equilibrium when considering fluctuations of quantities which characterize flows across a system, for instance work and particle currents. The most widely adopted method resorts to a two-point measurement (TPM) which consists in projectively measuring the observable of interest at two different times~\cite{Talkner2007,Esposito2009,Campisi2011}. The resulting probability distribution for changes in outcomes can then be used to recover, for instance, the celebrated work and current fluctuation relations~\cite{Crooks1999,Evans2002,Maes2003,Esposito2009,Talkner2007,Campisi2011,Hanggi2016}.

However, these projective measurements are challenging from an experimental point of view. To our knowledge, a direct implementation in a quantum system has only been realized in Ref.~\cite{Masuyama2018}.
This has led to a flurry of research to find alternative ways of measuring energy statistics. One famous approach is to perform Ramsey interferometry on the system \cite{Paternostro2013,Vedral2013} and has been successfully implemented in nuclear magnetic resonance experiments \cite{Serra2014}. 
Alternatively, one can use generalized quantum measurements, in which case the energy statistics can be inferred from a single measurement on an auxiliary system \cite{Paz2014,Chiara2015,Folman2017}. A different approach is based on dynamic Bayesian networks, used to study energy exchanges between two quantum systems that are initially correlated and locally thermal \cite{Micadei2020a,Micadei2021}.
Yet another approach exploits entanglement between a system and an ancilla, in such a way that the first of the two measurements is translated into a measurement of the entangled ancilla instead~\cite{Aguilar2022,Solfanelli2021}.

In this paper, we show that the energy statistics of a quantum system can be inferred from its collision with a particle prepared in a suitable initial state. Contrary to TPM and the alternatives mentioned above, our scattering setup is fully autonomous so that the colliding particle acts effectively both as an external driving and as an energy measurement device. Previous studies using a scattering approach have focused on identifying the conditions on the initial state of the incoming colliding particle that are needed to interpret the average energy exchanges in the system as heat or work \cite{Barra2021,Parrondo2022,Barra2022}. In this study, we identify the conditions for the state of the particle after the collision to encode the energy fluctuations that would be obtained from a two-point projective measurement of the system energy (often interpreted as fluctuating work when the system is initially in a canonical state). We find that this happens when the incoming colliding particle is semiclassical and narrow in energy compared to the system's energies, and we provide the relation between the time-dependent driving energy and the spatial potential energy of the scattering process. We also show that when its energy uncertainty is large, the change in kinetic energy due to the collision encodes the untouched work, i.e. the average energy change that would result from a non-autonomous driving without performing any projective measurement \cite{Hanggi2016}.

Scattering theory is used from high-energy to condensed matter physics to link quantum theory with experiments \cite{Razavy2003,Belkic2004,Taylor2006,Furrer2009}. 
It is also widely used to study open quantum systems \cite{Alicki1981,Alicki1983,Dumcke1985,Zeilinger2003,Hornberger2006,Hornberger2007,Hornberger2008,Hornberger2008}, quantum transport \cite{Landauer1957,Buttiker1992,Nazarov2009}, quantum thermodynamics \cite{vonOppen2018,Widera2021,Barra2021,Parrondo2022,Barra2022} and quantum metrology \cite{Widera2016,Schmidt2019,Widera2020}. 
Our results should therefore contribute to a better understanding of quantum fluctuations in a broad class of realistic systems.
As we shall see, since scattering maps are energy-preserving operations, our results can be used to provide realistic ways to implement otherwise highly idealized quantum resource theory operations \cite{Alhambra2016,Lostaglio2019,Gour2019}.



\emph{Two-point measurement scheme. ---} We consider a closed quantum system $S$, of dimension $N$, undergoing a time-dependent dynamics according to the Hamiltonian $H_S(t) =H_S + V(t) \nu$, where $H_S$ is the bare Hamiltonian, $V(t)$ is a non-vanishing function only in the interval $t \in (0,\tau)$ and $\nu$ is a time-independent operator. The bare energies of the system are defined by $H_S \ket{j}=e_j\ket{j}$, where $\{ \ket{j} \}_{j=1}^{N}$ is the basis of eigenvectors associated to its energy spectrum $\{ e_j \}_{j=1}^{N}$. We denote the energy gap (or Bohr transition frequency) between eigenstates $\ket{j}$ and $\ket{j'}$ by $\Delta_{j'j} \equiv e_{j'}-e_{j}$. 

In the interaction picture, the time-dependent protocol is implemented by a unitary operator $U_I(\tau)$ which obeys the Schr\"odinger equation in the interaction picture
   \begin{align}
    \label{vonneumann}
    \frac{d}{dt} U_I(t) = -\frac{i}{\hbar}V_I(t) U_I(t) \; ,
    \end{align}
with $U_I(0) = \mathbb{I}_S$ and $V_I(t) = V(t) e^{i H_S t /\hbar} \nu e^{-i H_S t /\hbar}$. According to the two-point measurement scheme, we perform two projective measurements in the eigenbasis of $H_S$, one before and one after the interaction takes place \cite{Talkner2007,Esposito2009,Campisi2011,Hanggi2016}. The energy distribution associated to this scheme is then
    \begin{align}
    \label{tpmdistribution}
    P^{\mathrm{TPM}}(W) = \sum_{j',j} \delta(W + \Delta_{j'j}) |\braket{j'|U_I(\tau)|j}|^2 \braket{j|\rho_S|j} \; ,
    \end{align}
where $\rho_S \equiv \rho_S(0)$ is the state of the system before the projective measurements are performed.
Although the last expression is valid for arbitrary energy spectrum, we consider without loss of generality that the spectrum is non-degenerate, i.e. we assume that $\Delta_{j'j} = 0$ implies $j'=j$. This is warranted since no energy statistics can be inferred within an energy-degenerate subspace, although the spectrum can be still have degenerate energy gaps. The average energy is the first moment of the distribution
    \begin{align}
    \label{worktpmdistribution}
    \langle W \rangle^{\mathrm{TPM}} & = \int dW P^{\mathrm{TPM}}(W) W \nonumber \\
    & = \mathrm{Tr}_S[H_S \rho_S]  - \mathrm{Tr}_S[H_S \rho^{\mathrm{TPM}}_S(\tau)]  \; .
    \end{align}
The second line expresses the average energy in a basis-independent fashion, where
    \begin{align}
    \label{tpmsystem}
    \rho^{\mathrm{TPM}}_S(\tau) = \sum_{j',j} \Pi_{j'} U_I(\tau) \Pi_{j} \rho_S \Pi_{j} U^{\dagger}_I(\tau) \Pi_{j'}  \;
    \end{align}
is the state of the system after the non-selective, two-point measurement scheme is performed and $\Pi_{j} \equiv \ket{j}\bra{j}$.

If the time-dependent protocol is implemented without measurements, then the probability distribution in Eq.~\eqref{tpmdistribution} cannot be defined. However, we can still define the average energy change in the system before and after the protocol
    \begin{align}
    \label{untouchedwork}
    \langle W \rangle^0 = \mathrm{Tr}_S[H_S \rho_S]  - \mathrm{Tr}_S[H_S U_I(\tau) \rho_S U^{\dagger}_I(\tau)] \; ,
    \end{align}
which is the so-called untouched work~\cite{Hanggi2016}. Note that Eqs.~\eqref{untouchedwork} and \eqref{worktpmdistribution} coincide if the initial state is diagonal in the energy eigenbasis \cite{Hanggi2016}.

\emph{Scattering setup. ---} We consider now a time-independent scattering process involving the same quantum system $S$ with bare Hamiltonian $H_S$. In addition, we also include the motional degrees of freedom $X$ of a particle of mass $m$ with kinetic energy operator $(\hat{p}^2/2m) \ket{p} = E_p \ket{p}$. Here $\{ \ket{p} \}$ are improper (non-normalizable) eigenstates whose position representation are plane waves $\braket{x|p} = \exp( \mathrm{i} p x / \hbar) / \sqrt{2 \pi \hbar}$ and $E_p = p^2 / 2m \geq 0$ is the kinetic energy. The total Hamiltonian reads
    \begin{align}
    H = H_0 + \mathcal{V}(x) = H_S \otimes \mathbb{I}_X + \mathbb{I}_S \otimes \hat{p}^2/2m + \mathcal{V}(x) \; ,
    \end{align}
where $\mathbb{I}_S$ and $\mathbb{I}_X$ are the identity operators on the Hilbert space of $S$ and $X$. The interaction operator is given by $\mathcal{V}(x) = \nu \otimes \lambda(x)$, where $ \lambda(x)$ is non-vanishing only inside the interval $x \in (0,a)$. In this scattering process, we picture the system $S$ as being at rest in the region $x \in (0,a)$, thus playing the role of a scatterer for the incoming particle.


We take the full system to be initially in a factorized state $\rho_S \otimes \rho_X$, with $\rho_X$ describing the initial state of the motion of the particle before the collision. According to scattering theory, the final state of the full system after a single collision
    \begin{align}
    \label{fullscattmap}
	\rho' = S \big( \rho_S \otimes \rho_{X} \big) S^{\dagger} \; ,
    \end{align}
where $S$ is the unitary scattering operator. Crucially, it is possible to show that $S$ is also energy-preserving $[S,H_0] = 0$ \cite{Taylor2006,Belkic2004,Barra2021}. From Eq.~\eqref{fullscattmap}, we can obtain the final state of motion of the particle
    \begin{align}
    \label{scattmapX}
	\rho'_X = \mathrm{Tr}_S[S \big( \rho_S \otimes \rho_{X} \big) S^{\dagger}] \; .
    \end{align}
In the same way, we can obtain the final state of the system $\rho'_S$ by performing the partial trace over $X$. 



We are now interested in obtaining the final kinetic energy distribution from Eq.~\eqref{scattmapX}. It can be formally expressed in terms of the final momentum distribution $\rho'_X(p) \equiv \braket{p|\rho'_X|p}$ as follows
    \begin{align}
    \label{kineticenergydistributionformal}
    \rho'_X(E_p) = \frac{m}{p(E_p)}\sum_{\alpha = \pm }\rho'_X(\alpha p(E_p)) \; ,
    \end{align}
where $p(E_p) = \sqrt{2m E_p}$ and the sum is over positive and negative momenta of the final distribution (see discussion below).

We obtain an explicit expression for $\rho'_X(p)$ from Eq.~\eqref{scattmapX} by expressing the scattering operator in the eigenbasis of $H_{0}$, denoted by $\ket{p,j}\equiv\ket{p}\otimes \ket{j}$
    \begin{align}
	\label{expresionS}
	\braket{p',j'|S|p,j}& =\frac{\sqrt{|pp'|}}{m}\delta(E_{p}-E_{p'}-\Delta_{j'j}) s_{j'j}^{(\alpha \beta)}(E) \; .
    \end{align}
This expression follows from energy conservation $[S,H_0]=0$ \cite{Taylor2006,Belkic2004,Barra2021}. In Eq.~\eqref{expresionS} $s_{j'j}^{(\alpha \beta)}(E)$ is the scattering matrix at total energy $E = E_p + e_j$ and $\alpha={\rm sign}(p')$ and $\beta={\rm sign}(p)$ accounts for the final and initial direction of the momenta,
which can be positive or negative. The pairs $(++),(+-),(-+),(--)$ correspond to transmission from the left, reflection from the left, reflection from the right and transmission from the right probability amplitudes, which can be obtained from the solutions of the stationary Schr\"{o}dinger equation~\cite{Landau1977,Sakurai1993}. Using Eq.~\eqref{expresionS} it can be shown that the unitarity of $S$ implies the unitarity of the scattering matrix $\sum_{j',\alpha} s^{(\alpha \beta)}_{j'j}(E) [s^{(\alpha \beta')}_{j'k}(E)]^* = \delta_{jk} \delta_{\beta \beta'}$ \cite{Barra2021}. We add that, strictly speaking, Eq.~\eqref{expresionS} implies that $s^{(\alpha \beta)}_{j'j}(E)$ is only defined if $E \geq \max{ \{ e_{j'},e_j \}}$, in which case we say that the channel $\ket{j} \rightarrow \ket{j'}$ is open \cite{Barra2021}. However, we can extend the scattering matrix to closed channels by defining $s^{(\alpha \beta)}_{j'j}(E) = \delta_{j'j} \delta_{\alpha \beta}$ if $E < \max{ \{ e_{j'},e_j \}}$ so that unitarity holds for both closed and open channels.

In this study, we consider that the initial state of motion $\rho_X$ is travelling to the right so that $\rho_X(p,p') \equiv \braket{p|\rho_X|p'}$ has support only in $p,p' \geq 0$ (a similar analysis holds for states travelling to the left). In practice, this happens if $p_0 \gg \sigma_p$ where $p_0 = \mathrm{Tr}_X[\hat{p} \rho_X]$ is the average momentum of the initial state and $\sigma_p^2 = \mathrm{Tr}_X[(\hat{p}^2 - p^2_0) \rho_X]$ is its variance. We then obtain an explicit form for Eq.~\eqref{kineticenergydistributionformal}
    \begin{align}
    \label{kineticenergydistribution}
    \rho'_X(E_p) & = \sum_{j',j,k} \braket{j|\rho_S|k} \rho_X(E_p + \Delta_{j'j},E_p + \Delta_{j'k}) \nonumber \\
    & \times \sum_{\alpha=\pm} s_{j'j}^{(\alpha +)}(E_p + e_{j'}) [s_{j' k}^{(\alpha +)}(E_{p} + e_{j'})]^* \; .
    \end{align}
In the last expression, the initial state of the particle is defined in terms of the kinetic energy
    \begin{align}\label{initialkineticenergydistribution}
    \rho_X(E_p,E_{p'}) = \frac{m}{\sqrt{p(E_p) p(E_{p'})}} \rho_X(p(E_p),p(E_{p'})) \; .
    \end{align}
Note that the diagonal elements $\rho_X(E_p) \equiv \rho_X(E_p,E_p)$ yield the initial kinetic energy distribution. The change in average kinetic energy due to the collision can now be retrieved from
    \begin{align}
    \label{kineticenergychange}
    \Delta E_X & = \int_0^{\infty} dE_p~E_p~[\rho'_X(E_p) - \rho_X(E_p)] \nonumber \\
    & = \mathrm{Tr}_S[H_S \rho_S] - \sum_{j',j,k} e_{j'}\braket{j|\rho_S|k} \int_{e_{j'}}^{\infty} dE~\rho_X(E - e_j,E - e_k) \nonumber \\
    & \times \sum_{\alpha=\pm} s_{j'j}^{(\alpha +)}(E)[s_{j' k}^{(\alpha +)}(E)]^* \; .
    \end{align}
The last expression follows from substituting Eq.~\eqref{kineticenergydistribution} in the first line of Eq.~\eqref{kineticenergychange} after some algebraic manipulation. Throughout the rest of the paper, we show how to connect Eq.~\eqref{kineticenergydistribution} and \eqref{kineticenergychange} of scattering theory to Eqs.~\eqref{tpmdistribution} and \eqref{worktpmdistribution} of the two-point measurement scheme. We then show how to recover the notion of untouched work and finally connect our results with quantum resource theories.


\emph{Connection to the two-point measurement scheme. ---} Under certain conditions, to be discussed below, Eq.~\eqref{kineticenergydistribution} can be written as 
    \begin{align}
    \label{result}
    \rho'_X(E_p) = \int dW P(E_p,W) \rho_X(E_p-W) \; ,
    \end{align} 
where
    \begin{align}
    \label{almostenergydistribution}
    P(E_p,W) \equiv \sum_{j',j} \delta(W + \Delta_{j'j}) P_{j'j}(E_p + e_{j'}) \braket{j|\rho_S|j} \; ,
    \end{align}
and $P_{j'j}(E) \equiv \sum_{\alpha} |s^{(\alpha+)}_{j'j}(E)|^2$ is the probability for a transition $\ket{j} \rightarrow \ket{j'}$ at total energy $E = E_p + e_{j'}$ due to the collision. The result in Eq.~\eqref{result} follows  from Eq.~\eqref{kineticenergydistribution} when the state of the particle is narrow in energy with respect to the system~\footnote{Alternatively, the result also follows if the system is initially diagonal $\braket{j|\rho_S|k} = \delta_{jk} \braket{j|\rho_S|j}$; however, in this paper we generally make no such assumption}, meaning
    \begin{align}
    \label{narrowenergy}
    \frac{|\Delta_{j'j}-\Delta_{k'k}|}{v_0} \gg \sigma_p \; ,
    \end{align}
for all non-degenerate energy gaps $\Delta_{j'j} \neq \Delta_{k'k}$, with $v_0 = p_0/m$ being the particle's average velocity. As studied in Refs.~\cite{Barra2021,Parrondo2022}, when condition \eqref{narrowenergy} is obeyed, the initial state appearing in Eq.~\eqref{kineticenergydistribution} can be written as $\rho_X(E_p + \Delta_{j'j},E_p + \Delta_{j'k}) \simeq \delta_{e_j,e_k} \rho_X(E_{p} + \Delta_{j'j})$. Thus, initial system coherences between non-degenerate states do not contribute to Eq.~\eqref{kineticenergydistribution} and Eq.~\eqref{result} follows directly. 


The function $P(E_p,W)$ defined in Eq.~\eqref{almostenergydistribution} differs from $P^{\mathrm{TPM}}(W)$ in Eq.~\eqref{tpmdistribution} in the transition probabilities. In the former, the transition probability $P_{j'j}(E_p + e_{j'})$ is independent of time but dependent on the energy of the system $S$, while in the latter $\braket{j'|U_I(\tau)|j}|^2$ is dependent on time but independent of the system's energy. However, as  shown in Ref.~\cite{Barra2022}, it is possible to simplify the scattering matrix when the kinetic energy --- or equivalently, the mass --- is the largest parameter involved, so that it depends only on the kinetic energy
   \begin{align}
   \label{scattmatrixsc}
    s_{j'j}^{(\alpha +)}(E_p + e_{j'}) \simeq \delta_{\alpha,+} \braket{j'|s(E_p)|j} \; ,
    \end{align}
where $s(E_p)$ acts only on $S$. This implies that $E_{p} \gg \lambda(x)$ and $p a_{\mathrm{min}} \gg \hbar$, where $a_{\mathrm{min}}$ is the minimum length scale over which the potential varies significantly. This ensures that the particle travels undisturbed through a given potential $\lambda(x)$ which can otherwise be arbitrary. Second, it implies that $E_{p} \gg \Delta_S$, so the kinetic energy is much larger than the maximum energy gaps of the system $\Delta_S \equiv \max (\Delta_{j'j})~\forall j',j$. 

The unitary operator $s(E_p)$, acting only on $S$, can be found by solving the Schr\"{o}dinger equation
    \begin{align}
    \label{unitary}
    \frac{\partial U_I(E_p,t)}{\partial t} = -\frac{i}{\hbar} V_I(E_p,t) U_I(E_p,t) \; ,
    \end{align}
with $U(E_p,-\infty) = \mathbb{I}_S$, $U(E_p,+\infty) = s(E_p)$, and  $V_I(E_p,t) \equiv \lambda(p(E_p) t/m)e^{i H_S t / \hbar} \nu e^{-i H_S t / \hbar}$. In other words, in the semiclassical limit a particle with kinetic energy $E_p$ induces an effective time-dependent interaction on the system $S$ equivalent to Eq.~\eqref{vonneumann}, provided we identify $V(t) \equiv \lambda(p(E_p)t/m)$. Such a time $t$ is related to the position $x$ at a fixed kinetic energy $E_p$ through the relation $t = m x /p(E_p)$, which can be used to write Eq.~\eqref{unitary} as an equivalent position-dependent equation. Note that since in our study $\lambda(x)$ is only non-zero inside an interval of length $a$, the previous relation automatically defines an interaction time $m a / p(E_p)$ at a fixed kinetic energy $E_p$. 

Finally, the condition $p_0 \gg \sigma_p$, already fulfilled for the states of motion considered, guarantees that $s(E_p)$ can be well approximated at $E_{p_0}$ \cite{Barra2022}. This means that $U_I(\tau) = s(E_{p_0})$, provided we identify $\tau \equiv a/v_0$ as the time of the collision. Therefore, in this regime Eq.~\eqref{almostenergydistribution} implies $P(E_p,W) \simeq P^{\mathrm{TPM}}(W)$. Substituting in Eq.~\eqref{result} we obtain our main result
     \begin{align}
    \label{resultsemiclassical}
    \rho'_X(E_p) = \int dW P^{\mathrm{TPM}}(W) \rho_X(E_p-W) \; .
    \end{align} 
The last expression shows that the final kinetic energy is the convolution of the two-point measurement distribution with the initial kinetic energy distribution. Moreover, applying the semiclassical regime to the integral in Eq.~\eqref{kineticenergychange} and taking initially diagonal states or narrow states obeying \eqref{narrowenergy}, we find $\Delta E_X = \langle W \rangle^{\mathrm{TPM}}$.


\emph{Connection to untouched work. ---} We now inquire what happens when the coherences are initially present and cannot be ignored, so that condition \eqref{narrowenergy} no longer holds. In particular, we look at states of motion which are broad in energy and narrow in position
    \begin{align}
    \label{broadenergy}
    \sigma_p \geq \frac{\hbar}{2 \sigma_x} \gg \frac{|\Delta_{j'j}-\Delta_{k'k}|}{v_0} \; ,
    \end{align}
where $\sigma_x = \mathrm{Tr}_X[(\hat{x}^2 - x^2_0) \rho_X]$ and $x_0 = \mathrm{Tr}_X[\hat{x} \rho_X]$ is the average position. As studied in Refs.~\cite{Barra2021}, the semiclassical regime together with conditions \eqref{broadenergy} and $\tau \ll \hbar / \Delta_S$ imply that $\rho_X(E_p + \Delta_{j'j},E_p + \Delta_{j'k}) \simeq \rho_X(E_{p})$. In this case, we find from Eq.~\eqref{kineticenergydistribution} that $\rho'_X(E_p) \simeq \rho_X(E_p)$, so that the final kinetic energy distribution is approximately unaffected by the collision with the system, carrying no information about its energy statistics. However, the second line of Eq.~\eqref{kineticenergychange} now yields $\Delta E_X = \langle W \rangle^{0}$ and we recover the untouched work \footnote{Note that we cannot use the first line of Eq. (13) to compute $\Delta E_X$ with the zero-order approximation $\rho_X'(E_p)\simeq \rho_X(E_p)$ for very broad states.}. 
We can thus interpolate between the notion of touched $\langle W \rangle^{\mathrm{TPM}}$ and untouched $\langle W \rangle^{\mathrm{0}}$ by varying the energy width of the particle scattering with the system.

\emph{Numerics. ---} We illustrate our results with  a $1/2$-spin system with Hamiltonian $H_{S} = (\Delta_S/2) \sigma_{z}$ and system interaction $\nu = J \sigma_x$, where $\Delta_S$ is the energy gap and $J \in \mathbb{R}$. We consider two different initial states for the system: a thermal state $\rho_S^{\beta} = \exp(-\beta H_S)/\mathrm{Tr}_S[\exp(-\beta H_S)]$ and a pure state with thermal populations $\rho_S^{\mathrm{coh}} = \rho_S^{\beta} + C$, where $C$ is only non-zero in the off-diagonal elements. The state of motion of the particle is modelled by a Gaussian wave packet $\rho_X(p,p') = \phi(p) \phi^*(p')$ where $\phi(p) = (2 \pi \sigma_p)^{-1/4} \exp[-(p-p_0)^2 / (4 \sigma^2_p)] \exp(-i p x_0 / \hbar)$ and the potential in space is given by a barrier $\lambda(x) = V_0$ for $x \in (0,a)$ and zero otherwise. The exact scattering matrix $s_{j'j}^{(\alpha \beta)}(E)$ is computed by solving the coupled and non-linear equations of multi-channel scattering theory \cite{Razavy2003}. We compare our results in the semiclassical regime with TPM by solving Eq.~\eqref{vonneumann} with a time-dependent potential $V(t) = V_0$ for $t \in (0,\tau)$ and zero otherwise, where $\tau = a/v_0$.

\begin{figure}[h!]
\centering
\begin{subfigure}
  \centering
  \includegraphics[width=7.7cm]{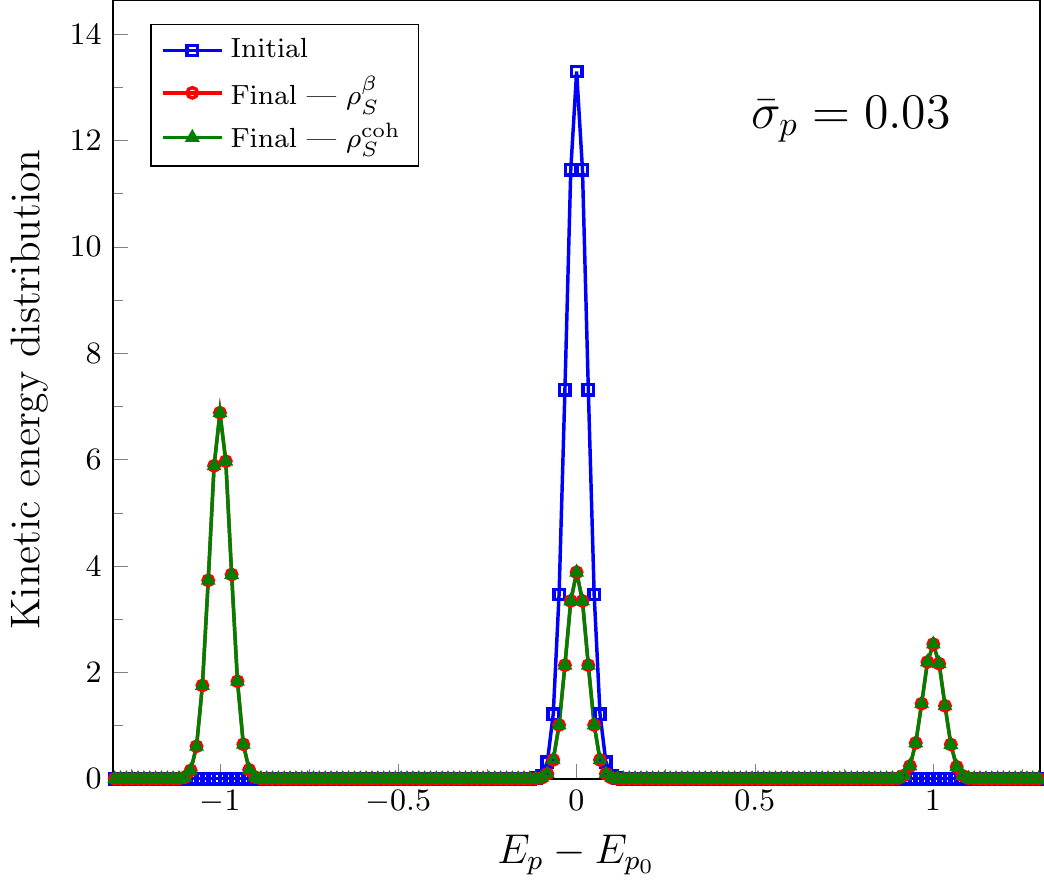}
\end{subfigure}
\begin{subfigure}
  \centering
  \includegraphics[width=7.7cm]{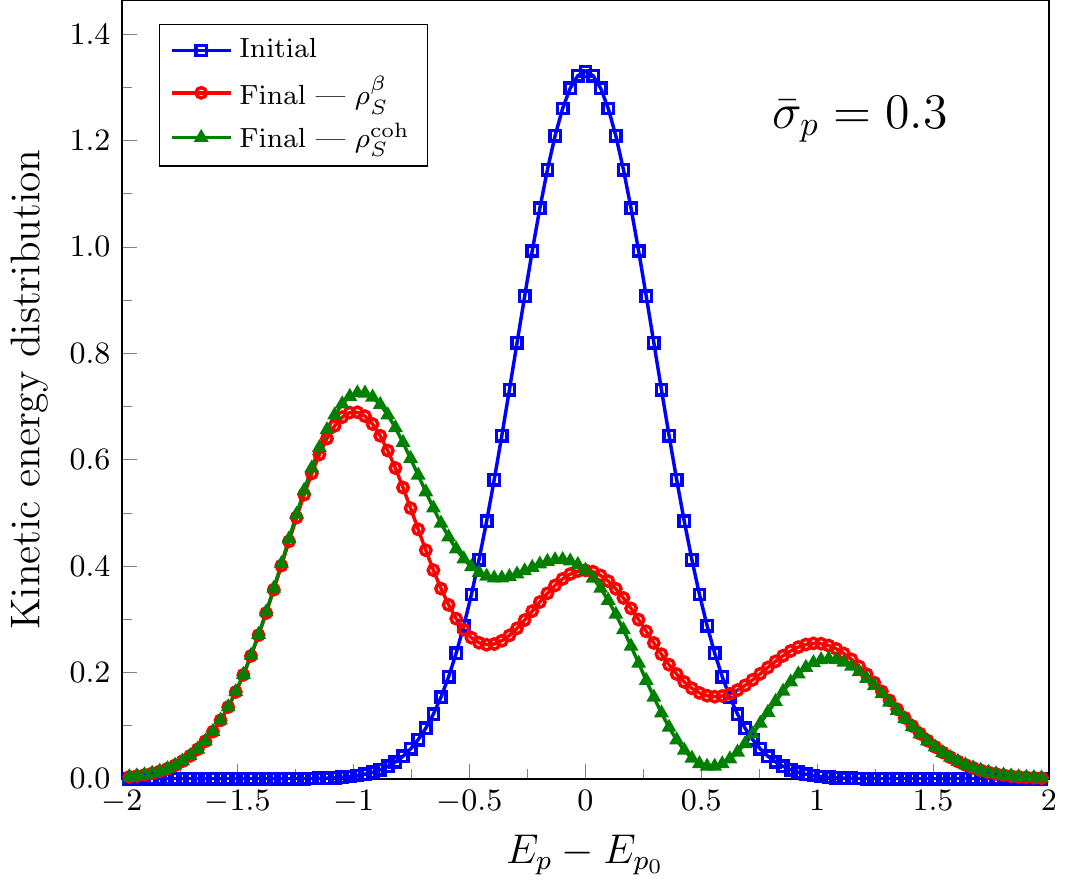}
\end{subfigure}
\begin{subfigure}
  \centering
  \includegraphics[width=8.05cm]{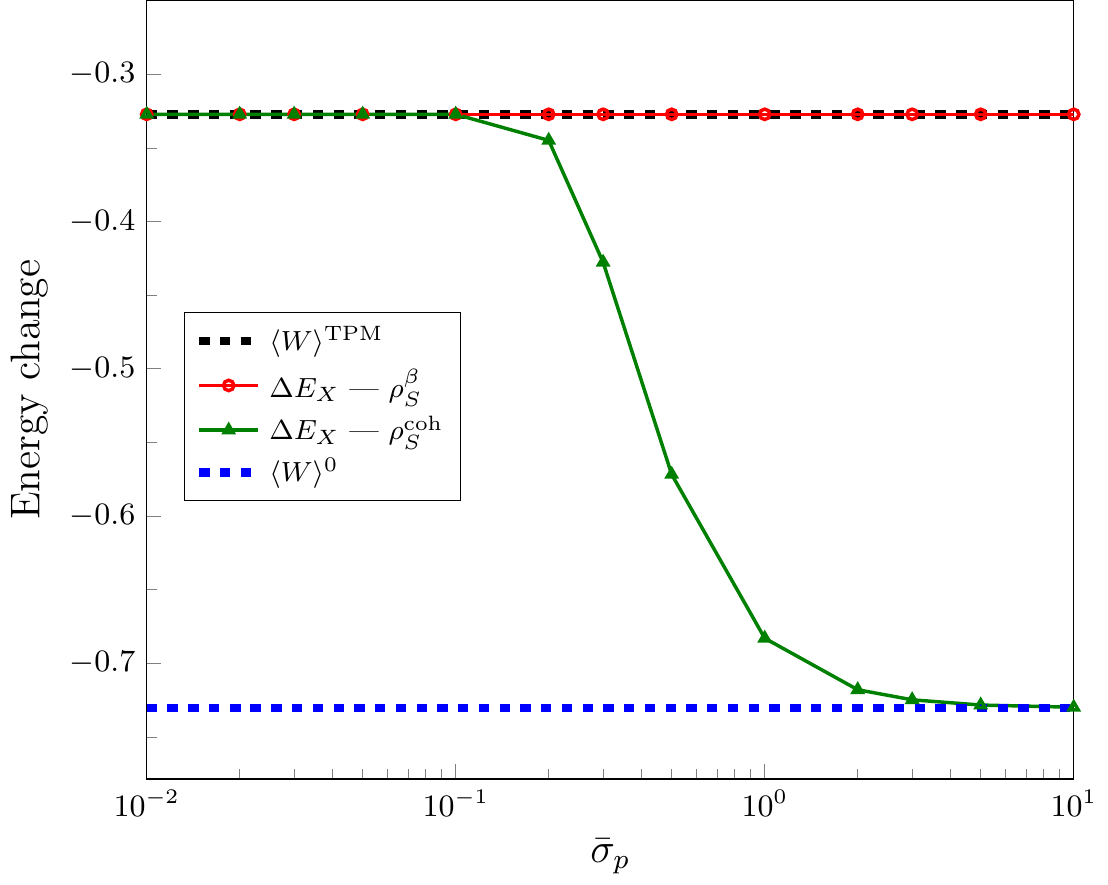}
\end{subfigure}
\caption{\label{fig:results} First and second panels: Initial and final kinetic energy distribution for different values of normalized momentum uncertainty $\bar{\sigma}_p \equiv \sigma_p v_0 / \Delta_{S}$ (computed from Eqs.~\eqref{initialkineticenergydistribution} and \eqref{kineticenergydistribution}, respectively). Third panel: Average change in kinetic energy as a function of $\bar{\sigma}_p$, along with the average energy from the TPM and untouched work (computed from Eqs.~\eqref{kineticenergychange}, \eqref{worktpmdistribution} and \eqref{untouchedwork}, respectively). Parameters: $\Delta_S = \beta = V_0 = J = 1$, $a = 1$, $v_0 = 1$ and $ m = 1000$.}
\end{figure}

In Fig.~\ref{fig:results}, we show the final kinetic energy distribution in Eq.~\eqref{kineticenergydistribution} and the initial distribution for different values of normalized momentum uncertainty $\bar{\sigma}_p \equiv \sigma_p v_0 / \Delta_S$ in the semiclassical regime. For narrow wave packets (first panel), we see that the final distribution yields well-defined and non-overlapping peaks corresponding to the different energy gaps of the system, independently of initial coherences. Accordingly, as we show in the third panel the average change in kinetic energy is equivalent to $\langle W \rangle^{\mathrm{TPM}}$. As we increase $\sigma_p$, the wave packet becomes broad in energy and the final kinetic energy distribution presents overlapping peaks whose amplitude depends on initial coherences (second panel). In the limit of very broad wave packets, the kinetic energy change approaches the untouched work $\langle W \rangle^{\mathrm{0}}$.

\emph{Connection to quantum resource theories. ---} We now connect our results to quantum resource theories \cite{Alhambra2016,Lostaglio2019,Gour2019,Landi2021}. In Ref.~\cite{Alhambra2016}, the system $S$ interacts with another quantum system $A$ whose Hamiltonian is $H_A = \gamma \hat{p}$, where $\gamma$ has units of velocity.
The unitary $U$ describes the evolution of $S$ and $A$ after they interacted
\begin{align}
    \label{fullsystemresource}
    \rho'_{SA} = U (\rho_S \otimes \rho_A) U^{\dagger} \; .
   \end{align}  
It is assumed to satisfy energy conservation $[U,H_S + H_A]=0$ and $[U,\hat{x}]=0$, implying that $U$ can be written as
    \begin{align}
    \label{unitaryresource}
    U = e^{-i H_S \otimes \hat{x} / (\hbar \gamma)} (V_S \otimes \mathbb{I}_A ) e^{i H_S \otimes \hat{x} / (\hbar \gamma)} \; ,
    \end{align}
where $V_S$ is an arbitrary unitary operator that acts on the system. A connection with the scattering setup is made by computing the final momentum distribution of $A$ given by $\rho'_{A}(p) = \braket{p|\rho'_A|p}$, which is obtained by tracing over $S$ in Eq.~\eqref{fullsystemresource} and using Eq.~\eqref{unitaryresource}, yielding
    \begin{align}
    \label{energydistributionresource}
    \rho'_A(p) & = \sum_{j',j,k} \braket{j|\rho_S|k} \rho_A(p + \Delta_{j'j}/\gamma,p + \Delta_{j'k}/\gamma) \nonumber \\
    & \times \braket{j'|V_S|j} \braket{j'|V_S|k}^{*} \; ,
    \end{align}
where $\rho_A(p,p') = (2 \pi \hbar)^{-1} \int dx' \int dx e^{-ipx/\hbar} e^{ip'x'/\hbar} \braket{x|\rho_A|x'}$ is the Fourier transform of system $A$ in position space $\braket{x|\rho_A|x'}$. Performing a change of variables from momentum to energy $E_A \equiv \gamma p$ in Eq.~\eqref{energydistributionresource}, we see that the resulting final energy distribution $\rho_A(E_A) \equiv 
\gamma^{-1} \rho_A(E_A/\gamma)$ has the same form as Eq.~\eqref{kineticenergydistribution} in the semiclassical regime (see condition \eqref{scattmatrixsc} and accompanying discussion). It follows that Eq.~\eqref{energydistributionresource} can be interpreted as the final kinetic energy distribution of a very massive particle $A$ after it collides with system $S$. Indeed, such an interpretation is corroborated by the fact that the Hamiltonian $H_A = \gamma \hat{p}$ has been used before to model very massive particles whose position in space can be used as a clock (see discussion after Eq.~\eqref{unitary} and Refs.~\cite{Aharonov1984,Aharonov1998,Woods2018}). As we have shown in this study, the energy statistics of $S$ can then be directly retrieved from Eq.~\eqref{energydistributionresource} when the system is initially diagonal or when the particle is narrow in energy, thus bypassing the need for a direct implementation of TPM.

\emph{Discussion and conclusion. ---} Inferring the energy statistics through semiclassical scattering with narrow energy states is an example of a ``which-path" experiment \cite{Scully1991,Zeilinger1996,Scully1997,Schlosshauer2007}. The colliding particle plays the dual role of inducing an energy jump in the system and carrying away ``which-jump" information in its kinetic energy distribution. However, acquiring ``which-jump" information automatically precludes the acquisition of information about coherences initially present in the system --- in this sense, information about ``which-jump" and information about superpositions are complementary \cite{Scully1991,Zeilinger1996}. On the opposite extreme, an incoming particle which is very broad in energy drives the system without carrying away information about it in its energy distribution. Accordingly, the average kinetic energy change corresponds to the untouched work, which depends on initial coherences in the system. We illustrated the transition between average work in the TPM scheme and the untouched work by changing the energy width of the incoming particle. 

In summary, we have shown that retrieving the energy statistics for a driven time-dependent isolated quantum system is possible by devising an appropriate scattering setup. In contrast to previous approaches, we provide a method to probe the energy statistics directly which can be experimentally realized in controlled scattering experiments e.g. in cavity quantum electrodynamics \cite{Haroche2001,Haroche2006} or ultracold atoms setups \cite{Widera2016,Widera2020,Widera2021}. In this way, scattering emerges as a versatile tool to study open quantum systems and quantum thermodynamics if one can control the state of the incoming particle.

\section*{Acknowledgments}

SLJ acknowledges the financial support from a Marie Skłodowska-Curie Fellowship (Grant No. 101103884).
GTL acknowledges the financial support of the S\~ao Paulo Funding Agency FAPESP (Grant No.~2019/14072-0.). 
F. B. thanks Fondecyt project 1191441 and ANID – 
Millennium Science Initiative Program-NCN19-170.

\bibliography{sam-library.bib}




\end{document}